\documentclass[aps,reprint,groupedaddress]{revtex4-2}

\usepackage{amsmath,amssymb}
\usepackage{amsfonts}
\usepackage[squaren,cdot]{SIunits}
\usepackage{graphicx}
\usepackage{xcolor}
\usepackage{epstopdf}

\DeclareGraphicsExtensions{.pdf,.jpg,.png,.mps,.tif,.eps}

\begin{document}

\title{Line tension in a thick soap film}
\author{Th\'eo Lenavetier}
\author{Ga\"elle Audéoud}
\author{Marion Berry}
\author{Ana\"\i s Gauthier}
\author{Rapha\"el Poryles}
\author{Corentin Tr\'egou\"et}
\author{Isabelle Cantat}

\affiliation{Univ Rennes, CNRS, IPR (Institut de Physique de Rennes) - UMR 6251, F- 35000 Rennes.}
\date{\today}

\begin{abstract}
The thickness of freshly made soap films is usually in the micron range, and interference colors make thickness fluctuations easily visible. Circular patterns of constant thickness are commonly observed, either a thin film disc in a thicker film or the reverse. In this Letter, we  evidence  the line tension at the origin of these circular patterns. Using a well controlled soap film preparation, we produce a piece of thin film surrounded by a thicker film. The thickness profile, measured with a spectral camera, leads to a  line tension of the order of $10^{-10}$ N which drives the relaxation of the thin film shape, initially very elongated, toward a circular shape.
 A balance between line tension and air friction leads to a quantitative prediction of the relaxation process. Such a line tension is expected to play a role in the production of marginal regeneration patches, involved in soap film drainage and stability.
\end{abstract}


\maketitle
The stability of liquid foams and soap bubbles is controlled by the evolution of liquid film thickness, induced by evaporation \cite{champougny18} and capillary and gravitational  drainage, until the film bursts. Drainage is associated with fast in-plane motion in films and thickness heterogeneities  \cite{mysels,hudales90,carrier02, lhuissier12, frostad16}, often spatially organized as discs of thin film embedded in a thicker film, or the reverse. 
In films less than 100 nm thick, both interfaces interact through short-range forces of various origins, resulting into a disjoining pressure.  Nonmonotonic variations of the disjoining pressure with the film thickness are known to induce a line tension along the boundary of film domains of different  thicknesses  \cite{defeijter72, joanny86}. This phenomenon has been  characterized for the transition between  a very thin suspended film and a meniscus \cite{platikanov80a}, or for the transition between two black films \cite{exerowa81}.

In this Letter, we show that the boundary between two domains of different thicknesses, both thicker than 100 nm, also generates a line tension, despite the negligible value of the disjoining pressure. In the transition between the two domains, the interface is slightly tilted and the excess area produces by this tilt, multiplied by the surface tension of the solution, is the excess energy at the origin of this capillary line tension, of purely geometric nature.

Marginal regeneration spontaneously generates film domains of different thicknesses \cite{mysels} and  
this line tension has already been assumed, qualitatively,  to play a role in such foam film instabilities \cite{shabalina19, tregouet21}.

To produce and evidence this original line tension, we prepare  an elongated pattern of  thin film surrounded by a thicker film and measure the relaxation of the  pattern  to a circular shape, under the effect of the line tension. Its value, deduced from the thickness profile we measure, is  of the order of $10^{-10}$ N and the relaxation lasts a few seconds, with velocities of the order of 10 mm/s.
The very low interfacial shear viscosity of our foaming solution  \cite{zell14}, rules it out as a significant friction mechanism.
 Considering the viscous friction of air only, and using the analytical prediction established in  \cite{saffman76,hughes81}, we are able to predict the relaxation rate as a function of the measured line tension, without adjustable parameter. This good agreement validates our line tension measurement. 
 
The line tension revealed by this work, and more generally the anisotropic interfacial stress induced by thickness gradients, whose tensor is given in this study, should therefore be taken into account in film drainage models, and potentially in experiments where foam films are used to investigate 2D turbulence \cite{kellay95, seychelles08}.

\begin{figure}
\centering
\includegraphics[width=0.48\textwidth]{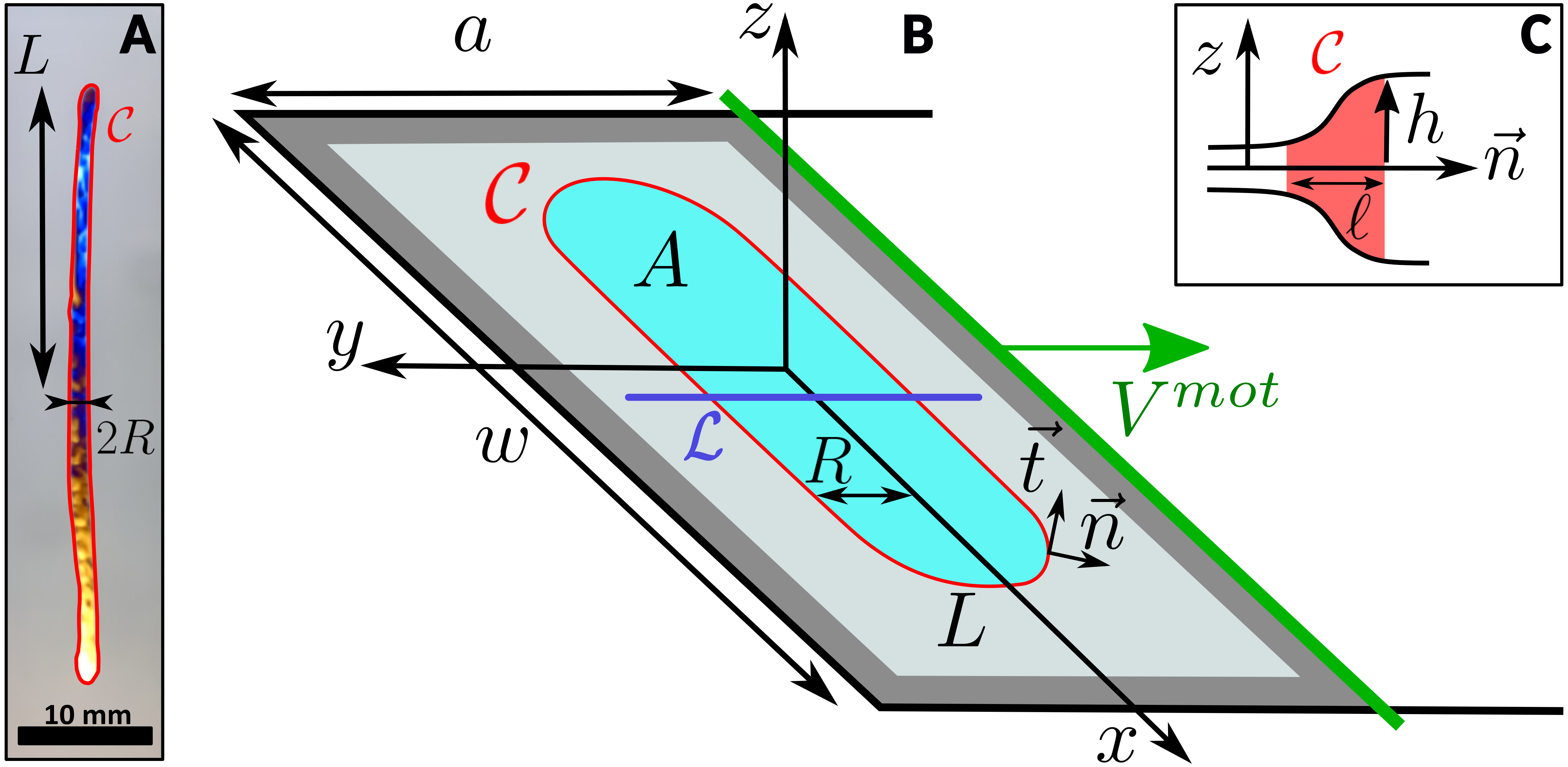}
\caption{Experimental setup and notations used in the text. (A) Image of the film recorded by the top camera. (B) Schematic view of the setup. The black and green thick lines represent  respectively the static and mobile edges of the frame. The thin film  (colored domain in (A), light blue in (B)) is separated from the light gray thick film by the red contour ${\cal C}$.  (C) Schematic thickness profile along $\mathcal{L}$, in the vicinity of  ${\cal C}$, on the right-hand side of Fig. 1B. 
\label{fig:setup}}
\end{figure}

\begin{figure}
\centering
\includegraphics[width=0.48\textwidth]{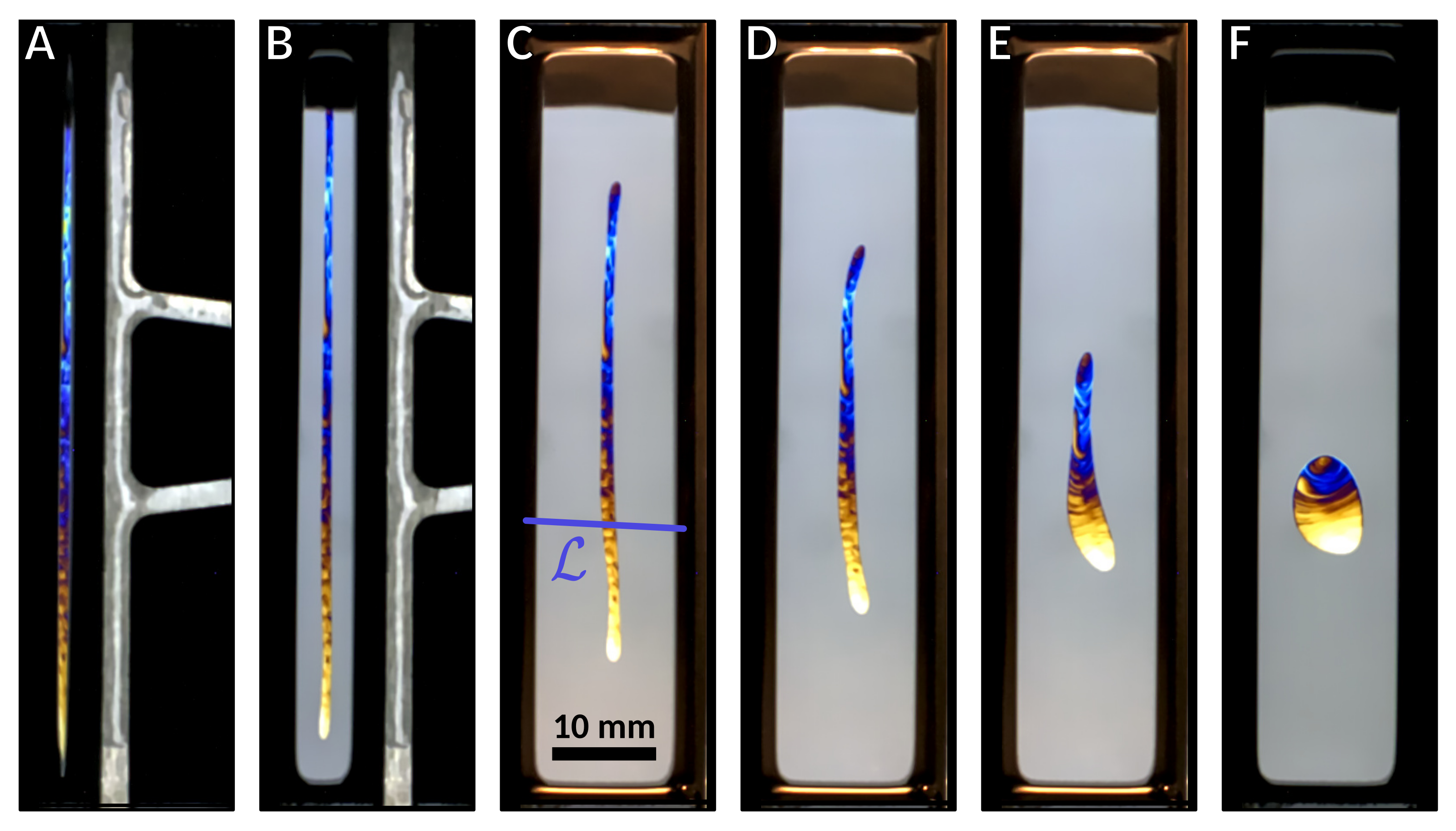}
\caption{
Top view of the film before, during and after deformation, at times $[-1,  -0.6, 0, 1, 3, 6 ]$ s. The colored central part is the thin film, the gray part is the thick film. The black boundary is the meniscus.
\label{fig:timelapse}}
\end{figure}

Using a deformable  horizontal frame of inner area $w \, a$, with $w= 62 $ mm and $a$ a variable width (see Fig. \ref{fig:setup}),  we produce a thickness pattern in a foam film. The $(x,y)$ plane is the midplane of the film.  
 We use a  mixture of sodium dodecyl sulfate (SDS, concentration 5.6 g/L, {\it i. e.} 2.4 CMC) and glycerol (15$\%$ in volume), of surface tension $\gamma_0 = 35$ mN/m (measured with the pendent drop method) and bulk viscosity $\eta_l=1.5$ mPa$\cdot$s. The interfacial shear viscosity  $\eta_s$ is shown to be below  $10^{-8}$ kg/s  in \cite{zell14}.
 
 Top views of the film, recorded with a color camera used at 30 frames per second, are shown in Fig.  \ref{fig:timelapse} at different times. The frame is first set at its smallest area ($a=2.1$ mm) and bathed in the foaming solution  to produce the thin part of the film (Fig. \ref{fig:timelapse}A), called the {\it thin film} hereafter. 
We let the film drain close to 3 min until its interference colors are mainly blue and yellow, indicating a thickness comprising between 100 and 300 nm. Then we move the mobile edge of the frame at a velocity $V^{mot}= 10$ mm/s during 1 s (Fig.%\ref{fig:timelapse}
1B) and a much thicker piece of  film, appearing gray,  is extracted from the meniscus surrounding the film  (hereafter,  the {\it thick film}).  The relaxation of the thin film toward a circular shape (Fig.\ref{fig:timelapse}
(C-F)) is studied after the motor stops, taken as time reference $t=0$.  The amplitude of the initial thickness fluctuations in the thin film is much smaller than the thickness difference between the thin film and the thick film and does not play any role in this relaxation.
The thickness profile of the transition between the thin and the thick parts of the film  is measured  with a hyperspectral camera (Resonon Pika L), at a rate of 50 frames per second, along a line $\cal{L}$ shown in Fig. \ref{fig:timelapse}C, as explained in \cite{bussonniere20}.

The boundary  ${\cal C}$ of the thin film  is  detected automatically and  characterized by its length $2L$ measured in the $x$ direction, its area $A$ and its width, defined as $2 R = A/(2L)$ (see Fig. \ref{fig:setup}A). At the beginning of the relaxation, the elongated shape is very regular and can be described as a rectangle $2L \times 2R$ with a hemidisc  of radius $R$ at both ends, with $R \ll L$.   At longer times, it becomes more fluctuating and a roughly circular shape is eventually  obtained at $t\approx 6$ s. Good reproducibility of the shape is obtained for $t<1.5$ s and the relaxation process it  quantitatively analyzed up to this time. 

\begin{figure}
\centering
\includegraphics[width=0.5\textwidth]{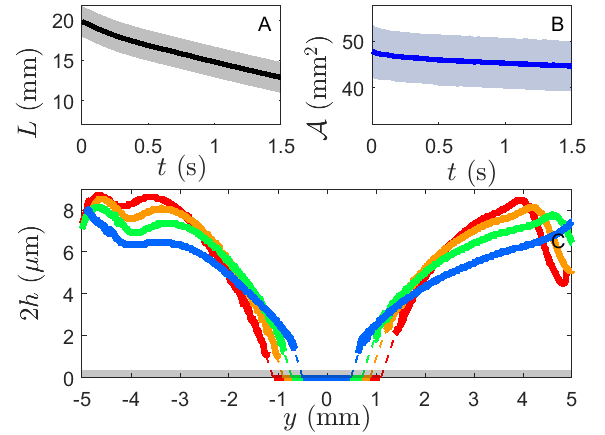}
\caption{Shape of the thin film as a function of time. (A) $L$ is half its diameter measured in the $x$ direction, and (B) $A$ is its area. As for the other figures, the data are averaged over 13 experiments, and the shaded area represents the standard deviation. 
(C) Example of thickness profiles at times 0 s (blue), 0.5 s (green), 1 s (yellow) and 1.5 s (red). 
The experimental resolution is indicated by the gray zone, and the width of the thin film is indicated by the continuous line at thickness 0.
 The dotted lines are parabolic interpolations between the measured domains. 
\label{fig:parameters}}
\end{figure}

The thin film area varies by at most 10$\%$ during the measurement time range (see Fig. \ref{fig:parameters}B). Moreover,  the whole thickness distribution in the thin film, indicated by the interference colors, remains qualitatively constant, thus excluding local compression or dilation in the thin film.

The film profiles are shown in Fig. \ref{fig:parameters}C, with $h(x,y)$ half the thickness of the film.
The  thickness of the thick and thin films are respectively of the order of 7 µm and below the resolution obtained with our current signal analysis, based on maxima detection in the spectrum of the light reflected by the film.
Near the thin film, the thickness profile is steep enough to blur the interference pattern and the thickness is also  not measurable. However, the  light patterns corresponding to the thin, flat film and to the steepest part of the thick  film clearly differ, and the boundary ${\cal C}$ between the two  is well defined, even if the thickness is not.
To reconstruct the missing part of the thick film profiles, we interpolate the thickness profile  by a parabola, imposing continuity of thickness and thickness derivative at the edge of the measured part of the thick film profile, and zero thickness at the boundary with the thin film. This choice of a parabola is arbitrary and other choices, e. g. a linear or order 3 interpolation, would lead to similar results. 

The  capillary forces governing the dynamics can be deduced from these profiles.
The thickness varies on characteristic horizontal distances of the order of $\ell \sim 1$ mm, yielding  $\nabla h \sim 10^{-2}$ (where $\nabla$ is the gradient operator in the (x,y) plane). In this small slope limit, the  pressure in the film, controlled by the Laplace pressure, scales as  $\gamma_0 h / \ell^2$ and the associated Poiseuille flow velocity scales as $(\gamma_0/\eta_l) h^3/\ell^3 \sim 10 \, $µm/s, which is negligible in the process. Moreover, as the flow occurs with negligible area variations (see Fig. \ref{fig:parameters}B),  we  assume incompressible interfaces, as classically made for spontaneous soap film dynamics \cite{mysels}.
 In this frame each elementary film element of volume $h dS$ is a closed system of constant thickness and constant area $dS$, moving at the uniform velocity ${\bf v}(x,y)$.  
We define  $\delta \gamma$ as  the difference between the local surface
tension $\gamma$ and the reference value $\gamma_0$, chosen in the middle of the film. This tension variation  $\delta \gamma$ ensures the constraint of incompressibility  $\nabla \cdot {\bf v}(x,y) =0$.

The 2D stress tensor acting on such film elements is computed in \cite{suppmat}, in the local basis ${\cal B}_e = ({\bf n}, {\bf t})$, defined in the ($x,y)$ plane so that $\nabla h= |\nabla h| {\bf n}$ (adapting to our specific case the general theory developed in \cite{edwards}). 
At order 2 in $\nabla h$, this tensor can be expressed as
$
\sigma_{cap}=
\sigma_{cap}^* + \sigma^f I 
$, 
with $I$ the identity matrix. The isotropic term is, using $\delta \gamma \ll \gamma_0$,
\begin{equation}
\sigma^f= 2 \left( \gamma_0 + \delta \gamma\right) +2  \gamma_0  h \Delta h     \, ,
\label{iso}
\end{equation}
where $\Delta$ indicates the 2D Laplacian operator.
The deviatoric part is, as  determined in \cite{suppmat},
\begin{equation}
\sigma_{cap}^* = \gamma_0
 \begin{pmatrix}
- (\nabla h)^2 & 0 \\
0 &  (\nabla h)^2 
\end{pmatrix}_{{\cal B}_e} 
\label{devia}
\end{equation}
and comes from the projection of the surface tension force  in the $(x,y)$ plane. 
The dominant term in the stress tensor is the surface tension $\gamma_0$ which is positive, thus indicating a traction. However, the contribution of $\sigma^*_{cap}$ shows that  this traction is slightly smaller  in the direction of the thickness gradient, and slightly larger in the perpendicular direction, which is at the origin of the line tension. 

The damping forces (per unit film area) are  the  friction on the gas phase $2 \, {\bf f}^g$ and on the surrounding film
$ \eta_f \Delta {\bf v}$ with $ \eta_f= 2 \eta_s + h \eta_l$ the film shear viscosity. It results from \cite{hughes81} that the air friction dominates if $\eta_f < 4 \times 10^{-8}$ kg/s, which is verified here, as $\eta_s < 10^{-8}$ kg/s and $ h \eta_l \sim 10^{-8}$ kg/s. 
However, $\eta_s$  depends on the foaming solution and may be much larger. In order to provide a general prediction, valid for a wide range of foaming solutions, we thus keep the air friction and the interfacial viscosity in the model.

The equation of motion is finally,  as already established  in \cite{bruinsma95} using another approach,
\begin{equation}
 2 \gamma_0 h \nabla (\Delta h)  + 2 \nabla \delta \gamma + 2 \eta_s \Delta {\bf v} + 2 {\bf f}^g = 0  \;  , 
\label{eqmo}
\end{equation}  
with the first two terms equal to $\nabla \cdot  \sigma_{cap}$, as derived in \cite{suppmat}.

As the capillary forces are localized along the thin film boundary ${\cal C}$, they can be interpreted as arising from a line tension, which considerably simplifies the problem. To this end,  we  define the local coordinates $(\xi, s)$, in the vicinity of  ${\cal C}$. The variable $\xi$ is zero on ${\cal C}$, and varies in the normal direction ${\bf n}$ whereas $s$ varies in the tangential direction ${\bf t}$. The definition of a line tension requires the localization condition $\ell \ll 1/\kappa$ with $\kappa$ the curvature of ${\cal C}$ : this is verified on the straight parts of ${\cal C}$, but not at the tips, where the curvature radius is $R \sim \ell$. For sake of generality, the tension is determined below for a generic curve ${\cal C}$ of small curvature, and will eventually be used in our case for the straight parts of ${\cal C}$ only.

The line tension is defined as the excess of capillary stress, with respect to the surface tension $\gamma_0$, integrated along a line perpendicular to the thickness transition.  It can thus be written as, for each interface,
\begin{equation}
 T = \frac{1}{2}\int_{\ell_{ -\infty}}^{\ell_\infty} {\bf t} \cdot (\sigma_{cap}- 2 \gamma_0 I) \cdot{\bf t} d\xi \, . 
 \label{tens}
\end{equation}
with $\ell_{-\infty}$ and $\ell_\infty$ the lower and upper bounds of the integration domain, larger than the transition width.

This integral depends on $h$ but also on $\delta \gamma$, which is determined below using Eq. \eqref{eqmo} in the domain $|\xi|  < \ell$. 
There, the first term in Eq. \eqref{eqmo} scales as $\gamma_0 h^2/\ell^3$. In the limit of small $\ell$, it is  much larger  than the viscous forces which vary smoothly across the transition domain.
 Eq. \eqref{eqmo} thus becomes $ \partial \delta \gamma / \partial \xi  = - \gamma_0 h \partial^3 h/\partial \xi^3$. 
By integration, we obtain  at first order in $\ell \kappa$ and for small $\xi$,

\begin{equation}
 \delta \gamma  = \gamma_0 \left ( \frac{1}{2}\left(\frac{\partial h}{\partial \xi}  \right)^2 - h \frac{\partial^2 h}{\partial \xi^2} \right ) \, . 
 \label{tensionloc}
\end{equation}
Inserting this expression into Eq. \eqref{tens} we obtain
%Finally the line tension is, expressed for a single interface,  
\begin{equation}
 T =  \gamma_0 \int_{\ell_{ -\infty}}^{\ell_\infty} \left(\frac{\partial h}{\partial \xi}  \right)^2 d\xi  \, . 
\label{eqtension}
\end{equation}
Note that the tension value is twice the energy excess per unit length of line associated to the thickness gradient.

\begin{figure}
\centering
\includegraphics[width=1\linewidth]{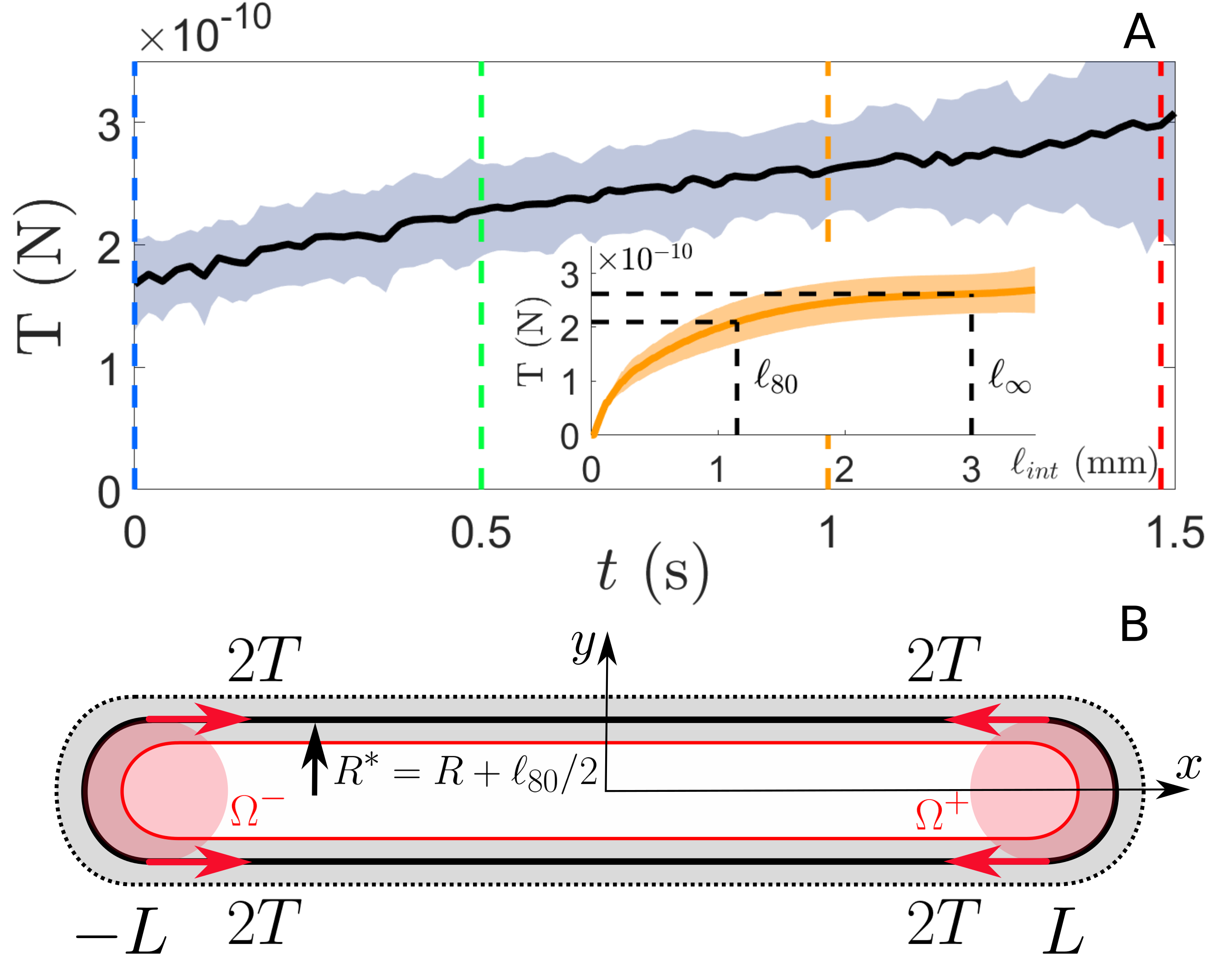}
\caption{(A) Line tension $T$ as a function of time, determined from the thickness profiles using Eq. \eqref{eqtension}, with  $\ell_{-\infty}= 0$,   $\ell_\infty= 3$ mm and $\xi =y$.  
The vertical dotted lines are color-matched in time with the thickness profiles of Fig. \ref{fig:parameters}C.
  Inset: partial values of the tension as a function of the integration upper bound $\ell_{int}$, obtained for the profile at $t=1$ s. The tension reaches $80 \%$ of its total value for $\ell_{int}= \ell_{80}$.
(B) Mapping of the observed flow on the problem solved in \cite{hughes81}. The thickness transition is shown in gray, with its center line ${\cal C}^*$ (bold black line), at the distance $\ell_{80}/2$ outside ${\cal C}$ (red line). The subdomains $\Omega^{\pm}$ are the red discs.     
\label{fig:T}}
\end{figure}
 
The experimental line tension values shown in Fig. \ref{fig:T} A,
have been determined with Eq. \eqref{eqtension}, using the experimental thickness profiles at each time, averaged over all the experiments, and with  the integration boundaries discussed below. 
As the angle between ${\bf n}$ and the $y$-direction is negligible, we have $\xi = \pm y$, respectively, for the left and right parts of the profile.
The thickness gradients are negligible in the thin film, so we impose  ${\ell_{ -\infty}} =0$. The relevant upper boundary is more difficult to chose: the inset of Fig. \ref{fig:T}A shows the partial tension values obtained when using an arbitrary upper integration  bound  $\ell_{int}$ in Eq. \eqref{eqtension} instead of $\ell_\infty$. 
A plateau value is obtained for $\ell_{int}$ between 2 and 3 mm, so we  choose $\ell_\infty = 3$ mm to define the experimental tension. Additionally, the width of the  transition domain is defined as $\ell_{80}(t)$,  the upper boundary value for which the tension reaches $80 \%$ of its total value. 

In this frame, the dynamical effects of  thickness variations are captured by a line tension acting along the line ${\cal C}^*$ located in the middle of the  transition domain at the distance $\ell_{80}/2$ from ${\cal C}$ (see Fig. \ref{fig:T}B). The resulting force (per unit line) acting on the film is $ - T \kappa^*(s) {\bf n} + (\partial T/\partial s) {\bf t}$, with $\kappa^*(s)$ the boundary curvature of ${\cal C}^*$.

The gradient $\partial T/\partial s$ is experimentally unknown. However, the thick film extraction velocity is very homogeneous all around the thin film, as well as  the meniscus size. We can thus safely assume that the initial transition profile, and consequently $T(t=0)$, does not depend on $s$. A strong assumption of the model is that the tension remains invariant at longer times.
In that case, the capillary force vanishes outside the region of the thin film tips and an analytical solution can be obtained.

To this end, we define around each thin film tip  a subdomain $\Omega^{\pm}$ limited by a flat cylinder,  centered at $ r_{M^{\pm}}= [\pm (L-R), 0]$ and of radius $R^*= R+ \ell_{80}/2$ so that the curved parts of ${\cal C }^*$ are in  $\Omega^{\pm}$ (see Fig. \ref{fig:T}B).
The deformations of the tips, associated to the increase of $R^*$ with time,  are much slower  than $dL/dt$ and
the whole subdomains $\Omega^\pm$ are moving at the uniform velocity  $\mp dL/dt$. 

Outside these domains Eq. \eqref{eqmo} becomes 
\begin{equation}
2 \nabla \delta \gamma + 2 \eta_s \Delta {\bf v} +  2 {\bf f}^g = 0 \; . 
\label{eqmo2}
\end{equation}  

The gas Reynolds number is of the order of $ R (dL/dt) \rho_g/ \eta_g \sim 1$, with $\rho_g= 1.2$ kg/m$^3$ and $\eta_g= 1.8 \, 10^{-5}$ kg/m/s   the density and viscosity of air. 
To obtain  an analytical prediction for the air damping force ${\bf f }^g$, we will neglect the air inertia, which  should be taken into account in a more refined model.

In this viscous limit, and if only  $\Omega^-$ moves, Eq. \eqref{eqmo2} can be  solved by a simple  mapping on the problem  solved in \cite{hughes81}, {\it i. e. } a flat cylinder translating in a viscous liquid membrane, as discussed in \cite{suppmat}.
The corresponding velocity field has been determined numerically in \cite{stone98}, and scales as  $dL/dt \;  R^*/r$ with $r$ the distance to the center of $\Omega^-$. As $R^*/(2L) \ll 1$ during the time range of measure, the velocity induced by the $\Omega^-$ motion  at the $\Omega^+$ position is negligible and the flow is thus  the superposition of the flows induced by the motion of each subdomain separately.
The force ${\bf F}_D$ acting  on the boundary of $\Omega^-$, due to the viscous friction of the gas phase and of the soap film, is determined in \cite{hughes81} and expressed as  ${\bf F}_D= \zeta  dL/dt \,  {\bf e}_x$, with $ \zeta$  a friction coefficient which depends on the Boussinesq number   $Bq =\eta_s/( R \eta_g)$ (see \cite{suppmat}).
 The force balance on the subdomain $\Omega^-$ involves this friction force ${\bf F}_D$ and the driving force $4T {\bf e}_x$ due to the 4 intersections between ${\cal C}^*$ and the boundary of   $\Omega^-$. This imposes  $ \zeta  dL/dt+ 4T=0$.

\begin{figure}
\centering
\includegraphics[width=0.45\textwidth]{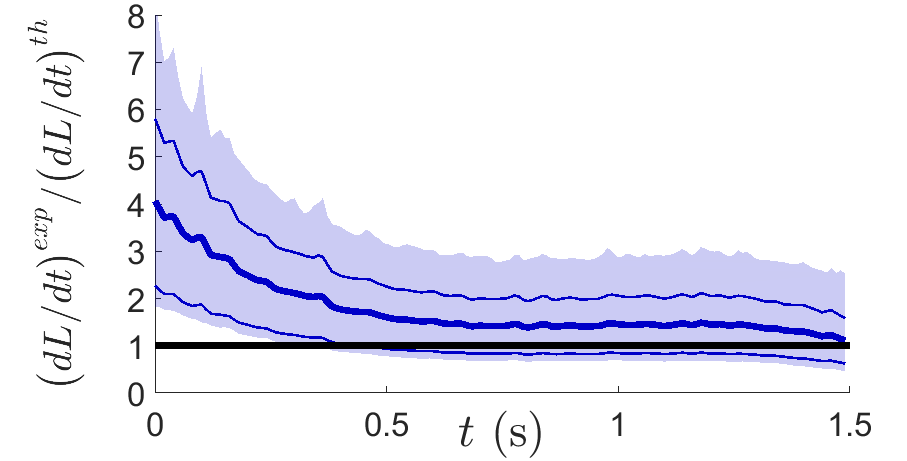}
\caption{Experimental value of the tip velocity $dL^{exp}/dt$, divided by its theoretical value given by Eq. \eqref{Ltheo}, in which $R^*=R+\ell_{80}/2$ (bold line), $R^*=R$ (bottom thin line) and $R^*=R+\ell_{80}$ (top thin line). Each curve is plotted with  a shaded area showing its standard deviation.
\label{fig:test_theo}}
\end{figure}

Assuming $\eta_s < 10^{-8}$ kg/s as measured  in \cite{zell14}, and using $R^* \sim 2$ mm   we find $Bq < 0.25$. In this low Boussinesq limit, the friction coefficient $\zeta$ reaches its asymptotic value  $\zeta_0= 16 \eta_g  R^* $ \cite{hughes81}, leading to 
\begin{equation}
\frac{dL}{dt} = - \frac{4 T}{16 \eta_g R^*}
\label{Ltheo}
\end{equation}
which involves only  experimentally known quantities.
The experimental relaxation velocity  
$dL^{exp}/dt$ is obtained  by differentiation of $L(t)$ shown in Fig. \ref{fig:parameters}A. Its theoretical value 
$dL^{th}/dt$, given by Eq. \eqref{Ltheo},  is obtained from the independent measurements of $T$ and $R^*$. The largest uncertainty arises from $R^*$, and is of the order of the width $\ell_{80}$  of the thickness transition domain. We thus plot in Fig. \ref{fig:test_theo} the predictions obtained with $R^*= R$ and $R^*= R+ \ell_{80}$.

We find that $(dL/dt)^{exp}/(dL/dt)^{th} = 1$  is  within our error bar for the time range $[0.5, 1.5]$, which  validates our line tension measurement, and its role as the driving force for the relaxation dynamics.
 At shorter times, the relaxation velocity is larger than predicted. A potential explanation for this reproducible deviation could be a residual air motion due the mobile edge, that may  last a fraction of second after the motor stop as $\textrm{Re}_{\textrm{gas}}\sim 1$. 

To conclude, this experiment quantifies the forces induced by thickness fluctuations, in a regime where disjoining pressure is negligible, and  shows that a  localized thickness gradient results into a line tension acting perpendicularly to the thickness gradient. As this tension is of purely geometrical nature, its expression Eq. \eqref{eqtension} should remain valid for nonhorizontal films, in the presence of gravity.
In foam films,  $z$-invariant in-plane motions occur with nearly no damping, and a tiny line tension, of the order of 0.1 nN for our thickness profile,  induces a thickness pattern relaxation toward a circular shape at a  velocity reaching 10 mm/s, only damped by the air friction. 
An extension of our analysis, based on the result of \cite{hughes81}, relates the pattern relaxation velocity to the value of the Boussinesq number.
We show in \cite{suppmat} that for foaming solutions having an interface viscosity above $4 \times 10^{-8}$ kg/s, the interface viscosity should be the  dominant damping factor. In that case,  a measure of a thickness pattern relaxation may  provide a measure of the interface viscosity, 
 which is an appealing application of our device.

\section*{Acknowledgments}
 This project has received funding from the European Research Council (ERC) under the European Union's Horizon 2020 research and innovation program (Grant Agreement No. 725094).

\end{document}

\title{SUPPLEMENTARY MATERIAL - Line tension in a thick soap film}
\author{Th\'eo Lenavetier}
\author{Gaëlle Audéoud}
\author{Marion Berry}
\author{Anaïs Gauthier}
\author{Rapha\"el Poryles}
\author{Corentin Tr\'egou\"et}
\author{Isabelle Cantat}

\affiliation{Univ Rennes, CNRS, IPR (Institut de Physique de Rennes) - UMR 6251, F- 35000 Rennes.}
\date{\today}

\maketitle

\subsection*{}
\section*{Interfacial stress tensor}

\begin{figure}
\centering
\includegraphics[width=0.3\textwidth]{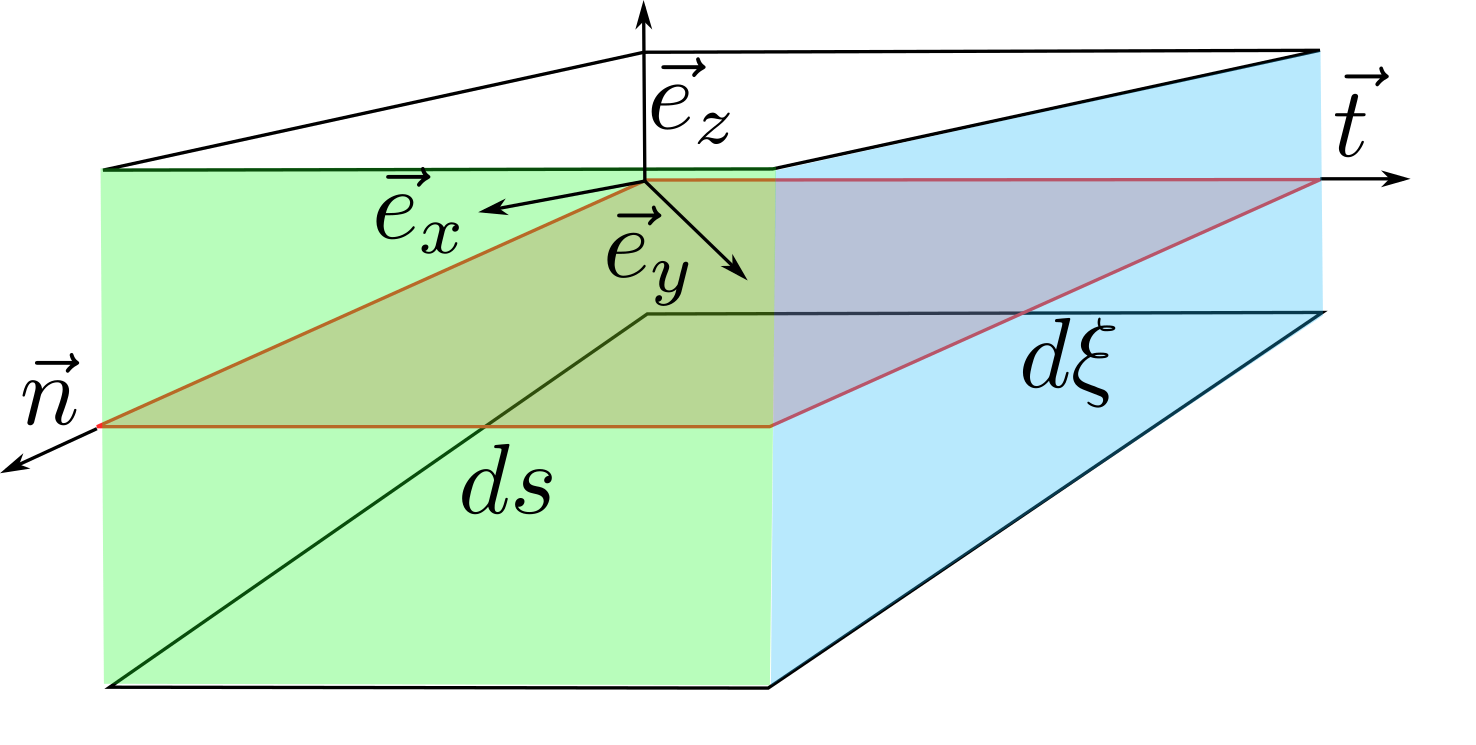}
\caption{Scheme of the film element $ds \, d\xi$ used for the determination of the stress tensor. The face 1, of normal $\vec{n}$, is in green, the face 2, of normal $\vec{t}$, is in blue. The thickness gradient is oriented along
$\vec{n}$.   
\label{fig:film_el}}
\end{figure}

We consider the piece of film represented in Fig. \ref{fig:film_el}. It is limited by the two interfaces and by the vertical planes which intersect the plane $z=0$ along the elementary lengths $d\xi \, \vec{n}$ and $ds \, \vec{t}$.    The unit vector  $\vec{n}$ is chosen along the thickness gradient and  the unit vector  $\vec{t}$ is perpendicular to  $\vec{n}$ (both are in the $(x,y)$ plane):
\begin{equation}
\vec{n} = (h_x, h_y, 0) / \sqrt{h_x^2 + h_y^2}   \quad \vec{t}= (-h_y, h_x, 0) / \sqrt{h_x^2 + h_y^2}  \, . 
\label{e1_e2}
\end{equation}

The 2D film stress tensor $\sigma_f$ can be built on this system, by considering the forces exerted on the lateral faces. Note that the air pressure is taken as the pressure reference, so that no forces are exerted on the top and bottom interfaces. The norm of the thickness gradient is a small parameter in the problem, as classically used in the lubrication approximation. To build the film stress tensor, we anticipate that  the forces exerted on the system are of order 2 in this parameter (as will be shown below), and we thus drop higher order terms.  Especially $\delta \gamma$ is of order 2. The differential operators $\nabla$ and $\Delta$ are the 2D gradient and laplacian in the $(x,y)$ plane.

Let us first consider the face 1 defined in Fig. \ref{fig:film_el}. It is a rectangle of  area $dS_1=2 h ds$ and of normal  $ \vec{n}$.
The pressure in the film is $P= - \gamma_0 \Delta h$, leading to the pressure force $d\vec{f}^P_1=2 \gamma_0  h  \Delta h\,  ds \, \vec{n}$. The local tension is $\gamma_0 + \delta \gamma$, acting on the length $ds$, at the top and bottom interfaces. The force orientation differs between both interfaces: it is along $\vec{n}^{\; top/bot} =  (\vec{n} \pm \| \nabla h \| \vec{e}_z)/ (1+ (\nabla h)^2)^{1/2}$. The resulting force is thus $d\vec{f}^T_1= 2 (\gamma_0 (1- (\nabla h)^2/2)  + \delta \gamma) ds  \vec{n}$. 

The face 2 is a trapezoid of area $dS_2= 2 h d\xi$ and of normal  $ \vec{t}$. The pressure force is  $d\vec{f}^P_2=2 \gamma_0  h  \Delta h \, d\xi \, \vec{t}$. The tension acts along the elementary length $ d\xi (1+ (\nabla h)^2)^{1/2}$, and the force is oriented along  $\vec{t}$, leading to the force  $d\vec{f}^T_2=2 (\gamma_0 (1+ (\nabla h)^2/2)  + \delta \gamma) d\xi  \vec{t}$. 

The interfacial stress tensor $\sigma_{cap}$ associated to these capillary forces verifies by definition  $d\vec{f}_1 =d\vec{f}^T_1+d\vec{f}^P_1 = \sigma_{cap} \cdot  \vec{n} ds$ and similarly  $d\vec{f}_2 = \sigma_{cap} \cdot  \vec{t} d\xi$. Its expression in the basis ${\cal B}_e = (\vec{n},  \vec{t})$ is thus 
\begin{equation}
\sigma_{cap} = \gamma_0
 \begin{pmatrix}
- (\nabla h)^2 & 0 \\
0 &  (\nabla h)^2 
\end{pmatrix}_{{\cal B}_e} + 2 \left( \gamma_0 (1+h \Delta h) + \delta \gamma  \right) I
\end{equation}

We thus get 
\begin{equation}
\sigma_{cap}=
\sigma_{cap}^* + \sigma^f I \, , 
\end{equation}
with the film tension $\sigma^f$ defined as the isotropic part of the interfacial stress
\begin{equation}
\sigma^f= 2 \left( \gamma_0 (1+ h \Delta h) + \delta \gamma  \right)  \, ,
\end{equation}
and $\sigma_{cap}^*$ defined as its deviatoric  part, which expression  in the initial basis ${\cal B}_0 =  (\vec{e}_x,  \vec{e}_y)$ is
\begin{equation}
\sigma_{cap}^*=
 - \gamma_0 \begin{pmatrix}
h_x ^2 - h_y^2 &  2 h_x h_y\\
2 h_x h_y &  h_y ^2 - h_x^2
\end{pmatrix}_{{\cal B}_0}  \, .
\end{equation}

The capillary force acting on the film element is $F_c=$ div $\sigma_{cap}$. We thus compute 

\begin{equation}
\mbox{div} (\sigma^f I)= 2 \gamma_0 h \nabla \Delta h  + 2 \gamma_0 \nabla h  \Delta h+ 2 \nabla (\delta \gamma)   \, ,
\end{equation} 
and 

\begin{align}
\mbox{div} \sigma_{cap}^* \cdot \vec{e}_x &= -\gamma_0 \left(2 h_x h_{xx} - 2 h_y  h_{xy} + 2 h_y h_{xy} + 2 h_x h_{yy} \right) \nonumber \\
& = -2 \gamma_0  h_x \Delta h  \nonumber\\
\mbox{div} \sigma_{cap}^* \cdot \vec{e}_y &= -\gamma_0 \left(2 h_x h_{xy} + 2 h_y  h_{xx} + 2 h_y h_{yy} - 2 h_x h_{xy} \right) \nonumber \\
& = -2 \gamma_0  h_y \Delta h  \nonumber\\
\mbox{div} \sigma_{cap}^* &= - 2  \gamma_0 \nabla h \Delta h
\end{align} 

Finally we obtain 
\begin{equation}
\mbox{div} \sigma_{cap}= 2 \gamma_0 h \nabla \Delta h + 2 \nabla (\delta \gamma)   \, ,
\end{equation} 
which is used to obtain eq. (3) in the main article.

\section*{Line tension}

The capillary forces are localized in the vicinity of a curve ${\cal C}$, in a domain of width $\ell$.
In the limit of small $\ell \kappa$, with $\kappa$ the curvature of ${\cal C}$, a line tension $T$ can be defined (for each interface). With this definition, $2T$  is the excess force exerted by one side of the film on the other one, across a line oriented along the normal to the boundary. This excess is considered with respect to the force field  far from $\cal C$, where the thickness is homogeneous and the surface tension equals to $\gamma_0$ at the dominant  order. 
This implies 
\begin{equation}
 T = \frac{1}{2}\int_{\ell_{ -\infty}}^{\ell_\infty} {\bf t} \cdot (\sigma_{cap}- 2 \gamma_0 I) \cdot{\bf t} d\xi \, . 
 \label{tens}
\end{equation}
A first step is to determine the surface tension within the transition domain. In this domain, the capillary forces dominate the viscous forces,  and thus control the surface  tension value.
In the small $\ell \kappa$ limit the force balance on a film element within the transition domain becomes  
\begin{equation}
 2 \gamma_0 h \nabla (\Delta h)  + 2 \nabla \delta \gamma = 0  \;  , 
\label{eqmoloc}
\end{equation}  
which can be simplified at leading order in $\ell \kappa$  into  
\begin{equation}
 \gamma_0 h \, \partial_{\xi \xi \xi} h  + \partial_\xi \delta \gamma = 0  \;  . 
\nonumber
\end{equation}  
%The force balance projected along $\vec{n}$ on the piece of film between $\ell_{-\infty}$ and $\xi$  (per unit length in the  $\vec{t}$ direction) is 
By integration between $\ell_{-\infty}$ and $\xi$ we get 
\begin{equation}
  \int_{\ell_{-\infty}}^\xi \left(\gamma_0 h \partial_{\xi \xi \xi} h   +  \partial_\xi \delta \gamma \right ) d\xi' = 0  \;  , 
\nonumber
\end{equation}
leading to, after integration by part,  
\begin{align}
\frac{\delta \gamma (\xi)}{\gamma_0} =&  \int_{\ell_{-\infty}}^\xi  \partial_\xi h \,  \partial_{\xi \xi} h \, d\xi'  - [ h \partial_{\xi \xi} h]_{\ell_{-\infty}}^\xi \nonumber \\
=& \frac{1}{2}\left(\frac{\partial h}{\partial \xi}  \right)^2 - h \frac{\partial^2 h}{\partial \xi^2} \; .
\label{eqmo}
\end{align}
We can now determine 
\begin{align}
&{\bf t} \cdot (\sigma_{cap}- 2 \gamma_0 I) \cdot{\bf t} = 2 \gamma_0 h \Delta h + 2 \delta \gamma + \gamma_0 (\nabla h)^2 \nonumber \\
&= \gamma_0 \left( 2 h \partial_{\xi \xi} h+ (\partial_\xi h)^2  - 2 h \partial_{\xi \xi} h + (\partial_\xi h)^2 \right)  \nonumber \\
&= 2 \gamma_0 (\partial_\xi h)^2   \; . 
\end{align}

Inserting this expression into eq. \eqref{tens} we obtain

%Finally the line tension is, expressed for a single interface,  

\begin{equation}
 T =  \gamma_0 \int_{\ell_{ -\infty}}^{\ell_\infty} \left(\frac{\partial h}{\partial \xi}  \right)^2 d\xi  \; . 
\label{eqtension}
\end{equation}

\section*{Viscosity upper bound}

\begin{figure}
\centering
\includegraphics[width=0.35\textwidth]{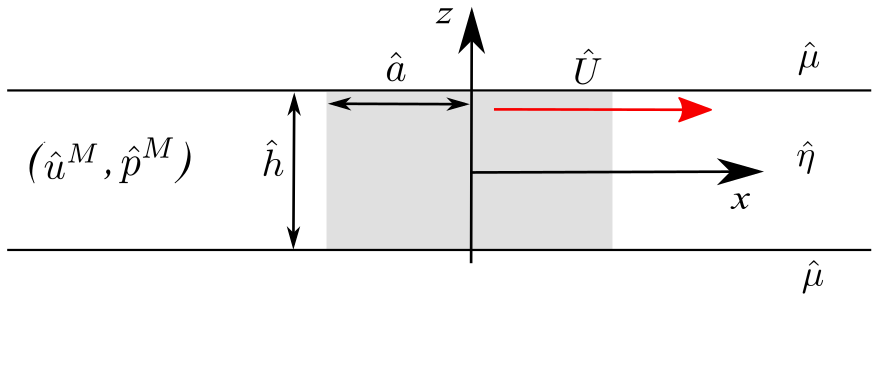}
\caption{Scheme of the problem solved in \cite{hughes81}. A cylindrical solid moves at the velocity  $\hat{U}$ in a liquid sheet (adapted from the Fig. 1 of \cite{hughes81}). 
\label{fig:schema_hughes}}
\end{figure}

The paper \cite{hughes81} addresses the problem of a disc of radius $\hat{a}$ and uniform thickness $\hat{h}$ moving at the velocity $\hat{U} \vec{e}_x$ in the plane $(x,y)$ of a  thin sheet of liquid of same thickness and viscosity $\hat{\eta}$ (the symbol $\hat{\cdot}$ indicates notations used in \cite{hughes81}, see Fig. \ref{fig:schema_hughes}). The bulk phases above and below this liquid sheet are fluids of viscosity $\hat{\mu}$.

In the viscous regime, the equation of motion of the thin sheet  is  (from eq. (2.18) of \cite{hughes81})
\begin{equation}
\hat{h} \hat{\eta} \nabla^2  {\bf \hat{u}^M} - \hat{h} \nabla \hat{p}^M +\hat{\bf F} = 0
 \end{equation}
with $\bf{\hat{u}^M}$ and $\hat{p}^M$ the velocity and the pressure of the sheet, which are assumed to be uniform across the film, and  ${\bf \hat{F}}$ the force exerted by the fluid phases above and below the sheet. These fluids verify the Stokes law. 

Using the transformations   $\bf{v}= \bf{\hat{u}^M}$, $2 \eta_s = \hat{h} \hat{\eta}$, $2 \delta \gamma = - \hat{h} \hat{p}^M$, $2 {\bf f}^g= 2 \hat{\bf F}$ we recover the equation (7) of our paper. 
With the assumption that  the whole domain $\Omega^-$ moves at the uniform velocity  $-dL/dt$, it can be identified with the solid disc, with   the transformation $-dL/dt = \hat{U}$ and $ R^*= \hat{a}$. 
The solution obtained in \cite{hughes81} for the total force exerted on the disc, due to the viscous friction of the top and bottom fluid phases and to the viscous thin sheet itself, can therefore be directly identified with the viscous forces acting on  $\Omega^-$. This force is given  in \cite{hughes81} as 
 a function of  $\hat{\varepsilon}= 1/Bq$, with  $Bq =\eta_s/( R^* \eta_g)$ a Boussinesq number.
 Using our notations, the result becomes
 ${\bf F}_D=  \zeta  dL/dt \,  {\bf e}_x$, with $ \zeta =  8 \pi \eta_g R^* \Lambda(Bq)$  a friction coefficient,  and $\Lambda(Bq)$ a semi-analytical function plotted in Fig. 2 of \cite{hughes81}.
 In the limit of small $Bq$ the friction coefficient is  $\zeta_0= 16 \eta_g  R^* $.  \\

We define our experimental friction $\zeta^{exp}$ as 
\begin{equation}
\zeta^{exp}  = - \frac{4 T}{(dL/dt)^{exp}}
\end{equation}
Comparing this value with the theoretical  friction $\zeta^{th}(Bq)$ we can determine an upper value for the interface viscosity of the film. 

\begin{figure}[!ht]
\centering
\includegraphics[width=0.4\textwidth]{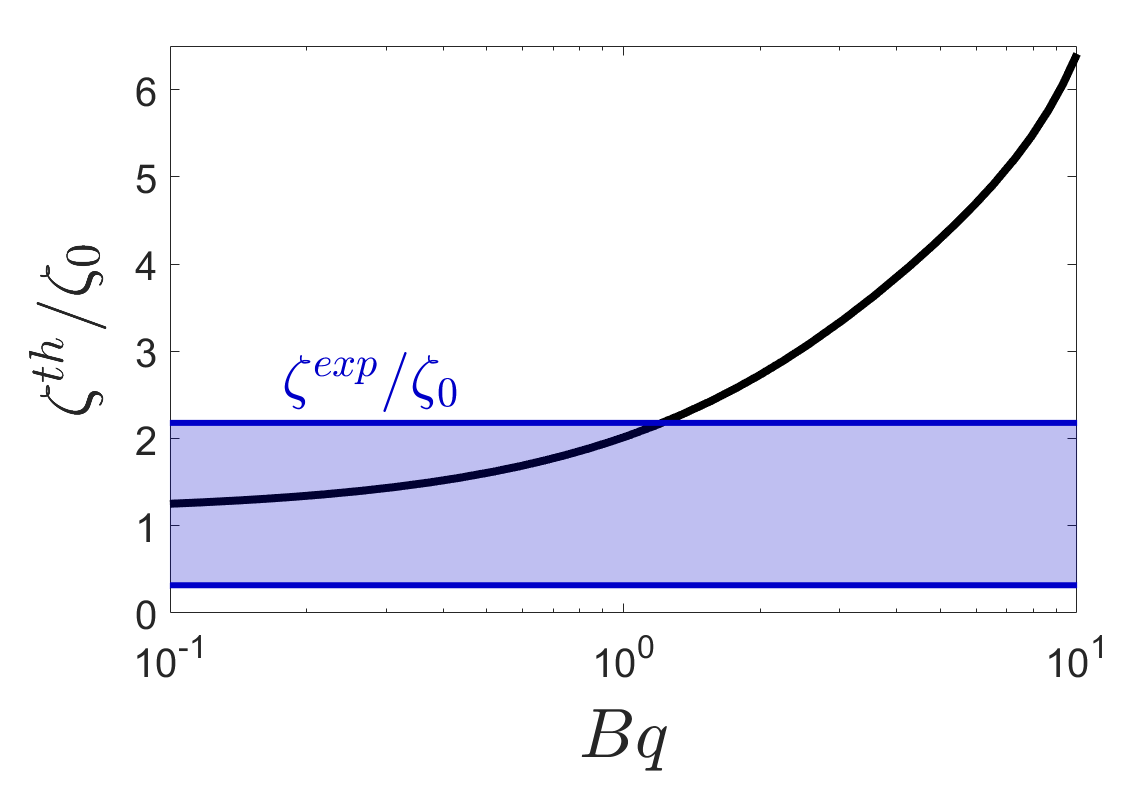}
\caption{Friction coefficient $\zeta$, renormalised by its value at low Boussinesq number. The black line is the  theoretical prediction obtained from the Fig. 2 in \cite{hughes81}. The blue domain is the value range compatible with the experimental values shown in the Fig 5 of the main paper, for the time range $[0.5 - 1.5]$ s. From the maximal acceptable value (the blue line at the top of blue domain) we deduce that $Bq <1$. 
\label{fig:viscomax}}
\end{figure}

In Fig. 5 of the main article, we plot the ratio
 \begin{equation}
\frac{\zeta^{exp}}{\zeta_0}  = - \frac{4 T}{16 \eta_g  R^* (dL/dt)^{exp}}= \frac{(dL/dt)^{th}}{(dL/dt)^{exp}}
\end{equation}
 as a function of time. A stable behavior is obtained for  $t \in [0.5, 1.5] \textrm{s}$ and, in this time range, the values compatible with our error bars verify  $\zeta^{exp}/\zeta_0  \in [0.3,  2.2]$. 
This acceptable range is shown in Fig. \ref{fig:viscomax} and compared to $\zeta^{th}(Bq)/\zeta_0 $. 
From the upper bound, we deduce that in our experimental condition the Boussinesq number is necessarily lower than 1. Using $R^* \sim 2$ mm, we obtain that  $\eta_s < 4 \times 10^{-8}$ kg/s, similar to the upper limit obtained in \cite{zell14}.

